\documentclass[reprint, amsmath, amssymb, floatfix, aps, superscriptaddress, nofootinbib, nobibnotes, onecolumn]{revtex4-1}
\usepackage{amsfonts}
\usepackage{mathrsfs}
\usepackage{natbib}
\usepackage{dblfnote}
\usepackage{graphicx}
\usepackage{hyperref}
\usepackage{bm}
\usepackage{amsmath}
\DeclareMathAlphabet{\mathscrbf}{OMS}{mdugm}{b}{n}
\usepackage{geometry}
\usepackage{indentfirst}
\usepackage{color}
\usepackage{acro}
\usepackage{svg}

\begin{document}

\title{Gravitational waves from primordial scalar and tensor perturbations}

\author{Zhe Chang} 
\affiliation{Institute of High Energy Physics, Chinese Academy of Sciences, Beijing 100049, China}
\affiliation{University of Chinese Academy of Sciences, Beijing 100049, China}

\author{Xukun Zhang}   
\affiliation{Institute of High Energy Physics, Chinese Academy of Sciences, Beijing 100049, China}
\affiliation{University of Chinese Academy of Sciences, Beijing 100049, China}

\author{Jing-Zhi Zhou} 
\email{zhoujingzhi@ihep.ac.cn}
\affiliation{Institute of High Energy Physics, Chinese Academy of Sciences, Beijing 100049, China}
\affiliation{University of Chinese Academy of Sciences, Beijing 100049, China}

\begin{abstract}
    We investigate the second order gravitational waves induced by the primordial scalar and tensor perturbations during radiation-dominated era. The explicit expressions of the power spectra of the second order GWs are presented. We calculate the energy density spectra of the second order GWs for the monochromatic primordial power spectra $\mathcal{P}_{\zeta}=A_{\zeta}k_*\delta\left( k-k_* \right)$ and $ \mathcal{P}_{h}=A_{h}k_*\delta\left( k-k_* \right)$. For large $k$ $\left( k>k_* \right)$, the primordial tensor perturbation will affect the total power spectrum of second order GWs significantly. For $f_*=1.3\times 10^{-3}$ and $A_{\zeta}=0.02$, the effects of the primordial tensor perturbation will lead to an around $100\%$ increase of the signal-to-noise ratio (SNR) for LISA observations if the tensor-to-scalar ratio $r=A_h/A_{\zeta}>0.4$.
\end{abstract}

\maketitle

\section{Introduction}\label{sec:1}
The Inflationary cosmology suggests that cosmological perturbations are originated from quantum fluctuations in early universe \cite{Sato:2015dga}.  Since the gravitational waves (GWs) were detected by LIGO and Virgo \cite{LIGOScientific:2016aoc}, GWs have attracted a lot of attention in cosmology. The further observations on stochastic GWs background could help us test inflationary models on small scale \cite{Inomata:2018epa}. 

The cosmological perturbations can be decomposed as scalar, vector, and tensor perturbations based on the behavior under the spatial coordinate transformation. The cosmological perturbations generated at inflation epochs are known as primordial perturbations. The power spectrum of primordial perturbations is one of the most important predictions in inflation theory. On large scales ($\gtrsim$1 Mpc),  the primordial power spectra are well constricted by the observations of the cosmic microwave background and large-scale structure \cite{Planck:2018vyg,Abdalla:2022yfr}. For the primordial scalar and tensor perturbations, it indicates a nearly scale-invariant primordial power spectrum of scalar perturbations with amplitude $\sim 2\times 10^{-9}$, and tensor perturbations with tensor-to-scalar ratio $r<0.06$ on large scales. However, the constraints on primordial scalar and tensor perturbations are much weaker on small scales ($\lesssim$1 Mpc) \cite{Bringmann:2011ut}. In recent years, scalar perturbations on small scales have been attracting a lot of interests. The primordial scalar perturbations with large amplitude on small scales have close relations with the primordial black holes (PBHs) and the scalar induced gravitational waves (SIGWs)  \cite{Mollerach:2003nq,Ananda:2006af,Baumann:2007zm,Saito:2008jc,Bugaev:2009zh,Saito:2009jt,Bugaev:2010bb,Nakama:2016gzw,Garcia-Bellido:2017aan,Kohri:2018awv,Cai:2018dig,Gong:2019mui,Carr:2020gox,Chang:2020tji,Zhou:2021vcw,Romero-Rodriguez:2021aws,Adshead:2021hnm,Domenech:2021ztg,Zhang:2022dgx,Chang:2022dhh}. Similarly, the primordial tensor perturbations on small scales could be much larger than it is on the large scales \cite{Nakama:2016enz,Nakama:2015nea}. The higher order GWs can also be induced by the primordial tensor perturbations.

The large primordial tensor perturbations on small scales may be realized by many models of early universe, such as $G^2$-inflation \cite{Kobayashi:2011nu}, cyclic/ekpyrotic models\cite{Boyle:2003km}, loop quantum cosmology \cite{Copeland:2008kz} and so on \cite{Nakama:2016enz}. In this paper, we study the second order GWs induced by the primordial scalar and tensor perturbations during a radiation-dominated (RD) era. In this case, these are three kinds of source terms for the second order GWs. Namely, the source term of the first order scalar perturbation $S\sim \phi^{(1)}\phi^{(1)}$, the source term of the first order scalar perturbation and the first order tensor perturbation $S\sim \phi^{(1)}h^{(1)}$, and the source terms of the first order tensor perturbation $S\sim h^{(1)}h^{(1)}$. These source terms and the corresponding kernel functions are studied systematically in this paper. Here, we consider a monochromatic power spectra for the primordial scalar and tensor perturbations with different tensor-to-scalar ratio $r=A_{h}/A_{\zeta}$ on small scales. We derive the explicit expressions of the power spectra of the second order GWs and calculate the corresponding energy density spectrum. The results show that the effects of the first order tensor perturbation enhance the density spectrum significantly for high frequency second order GWs.  

This paper is organized as follows. In Sec.~\ref{sec:2}, we present the equation of motion of the second order GWs. We calculate the kernel functions of seven source terms. In Sec.~\ref{sec:3}, the explicit expressions of the power spectra of the second order GWs are presented. We calculate the energy density soectrum of the second order GWs for a monochromatic primordial power spectra. The SNR LISA observations are calculated in this section.  Finally, the conclusions and discussions are summarized in Sec.~\ref{sec:4}.

\section{Equation of motion and kernel functions}\label{sec:2}
\begin{figure}
	\includegraphics[scale=0.8]{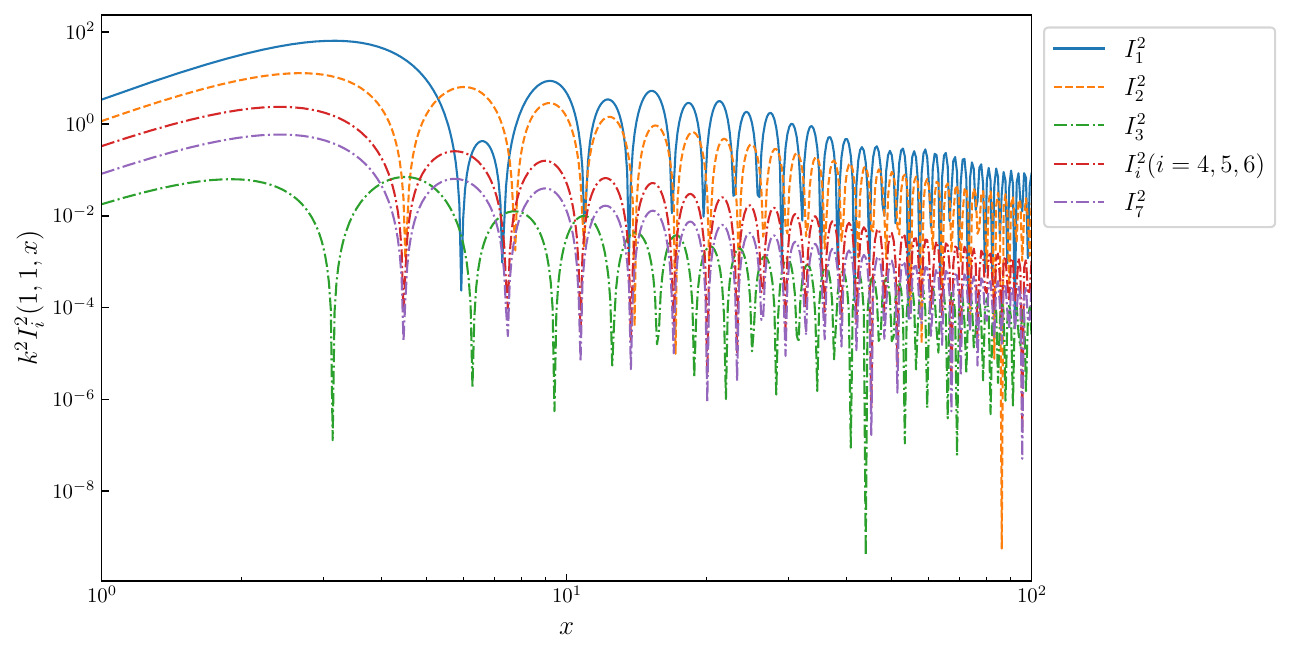}
	\caption{The kernel functions $I_i\ (i=1,2,...,7)$ calculated from Eq.~\ref{eq:I}. The kernel function of second order SIGW $I_1^{(2)}$ is much larger than other kernel functions. The second largest kernel function is $I_2^{(2)}$, which is the kernel function of the source term $S^{(2),\lambda}_2\sim \phi^{(1)}h^{(1),\lambda}$.}\label{fig:kernel_function}
\end{figure}

The perturbed metric in the flat FRW spacetime with Newtonian gauge is given by
\begin{equation}
	\begin{aligned}
		\mathrm{d} s^{2}&=a^{2}\Bigg[-\left(1+2 \phi^{(1)}\right) \mathrm{d} \eta^{2}+\left(\left(1-2 \psi^{(1)}\right) \delta_{i j}+h^{(1)}_{ij}+\frac{1}{2}h^{(2)}_{ij}\right)\mathrm{d} x^{i} \mathrm{d} x^{j}\Bigg] \ ,
	\end{aligned}
\end{equation}
where $\phi^{(1)}$ and $\psi^{(1)}$ are first order scalar perturbations, $h^{(n)}_{ij}$$\left( n=1,2 \right)$ are $n$-order tensor perturbations. The solutions of first order scalar perturbations and the first order tensor perturbation in momentum space are given by
\begin{equation}
	\psi(\eta,\mathbf{k}) = \phi(\eta,\mathbf{k}) = \Phi_{\mathbf{k}} T_\phi(k \eta)=\frac{2}{3}\zeta_{\mathbf{k}} T_\phi(k \eta) \ , \
	h^{\lambda,(1)}(\eta,\mathbf{k}) = \mathbf{h}^{\lambda}_{\mathbf{k}} T_{h}(k \eta) \ ,
\end{equation}
where $\zeta_{\mathbf{k}}$ and $\mathbf{h}^{\lambda}_{\mathbf{k}}$ are the primordial scalar and tensor perturbations respectively. The transfer functions $ T_\phi(k \eta)$ and $T_{h}(k \eta)$ in the RD era are given by \cite{Kohri:2018awv}
\begin{equation}\label{eq:T}
	T_{\phi}(x)=\frac{9}{x^{2}}\left(\frac{\sqrt{3}}{x} \sin \left(\frac{x}{\sqrt{3}}\right)-\cos \left(\frac{x}{\sqrt{3}}\right)\right) \ , \  T_{h}=\frac{\sin x}{x} \ ,
\end{equation}
where we have defined $x\equiv k\eta$. We use the $\texttt{xPand}$ package to study the higher order perturbations of Einstein equation on FRW spacetime. The equation of motion of second order GWs takes the form of 
\begin{equation}\label{eq:eq}
	\begin{aligned}
		h_{ij}^{(2)''}(\eta,\mathbf{x})+2 \mathcal{H} h_{ij}^{(2)'}(\eta,\mathbf{x})-\Delta h_{ij}^{(2)}(\eta,\mathbf{x}) =-4 \Lambda_{ij}^{lm} \mathcal{S}^{(2)}_{lm}(\eta,\mathbf{x}) \ ,
	\end{aligned}
\end{equation}
where $\mathcal{H}=a'/a$ is the conformal Hubble parameter, $\Lambda_{ij}^{lm}$ is the decomposed operator to extract the transverse and traceless terms. The source terms of the second order GWs are given by
\begin{equation}\label{eq:0}
	\begin{aligned}
		\mathcal{S}^{(2)}_{lm}(\eta,\mathbf{x})=\sum_{i=1}^{7}\mathcal{S}^{(2)}_{lm,i}(\eta,\mathbf{x})  \ ,
	\end{aligned}
\end{equation}
where
	\begin{eqnarray}
	 \mathcal{S}^{(2)}_{lm,1}(\eta,\mathbf{x})&=&\partial_{l} \phi^{(1)} \partial_{m} \phi^{(1)}-\frac{1}{ \mathcal{H}}\left(\partial_{l} \phi^{(1)'} \partial_{m} \phi^{(1)}+\partial_{l} \phi^{(1)}  \partial_{m} \phi^{(1)'}\right)+4 \phi^{(1)} \partial_{l} \partial_{m} \phi^{(1)}-\frac{1}{ \mathcal{H}^{2}} \partial_{l} \phi^{(1)'} \partial_{m}  \phi^{(1)'} \ ,
	\label{eq:1} \\
	\mathcal{S}^{(2)}_{lm,2}(\eta,\mathbf{x})&=& 10\mathcal{H}h^{(1)}_{lm}\phi^{(1)'}+3h^{(1)}_{lm}\phi^{(1)''}-\frac{5}{3}h^{(1)}_{lm}\Delta\phi^{(1)}-2\partial_{b}h^{(1)}_{lm}\partial^{b}\phi^{(1)}-2\phi^{(1)}\Delta h^{(1)}_{lm} \ ,
	\label{eq:2} \\
		 \mathcal{S}^{(2)}_{lm,3}(\eta,\mathbf{x})&=&-\frac{1}{2} h^{b,(1)'}_{l} h^{(1)'}_{mb} +\frac{1}{2}\partial_c h^{(1)}_{mb}\partial^c h^{b,(1)}_{l} \ ,
		 \label{eq:3} \\
		 \mathcal{S}^{(2)}_{lm,4}(\eta,\mathbf{x})&=&-\frac{1}{2} h^{bc,(1)}\partial_c \partial_l h^{(1)}_{mb}-\frac{1}{2} h^{bc,(1)}\partial_c \partial_m h^{(1)}_{lb}  \ , 
		 \label{eq:4} \\
		\mathcal{S}^{(2)}_{lm,5}(\eta,\mathbf{x})&=&-\frac{1}{2}\partial_b h^{(1)}_{mc}\partial^c h^{b,(1)}_{l}  \ ,  
		\label{eq:5}\\
		 \mathcal{S}^{(2)}_{lm,6}(\eta,\mathbf{x})&=&\frac{1}{2} h^{bc,(1)}\partial_c \partial_b h^{(1)}_{lm} \ , 
		 \label{eq:6} \\
		  \mathcal{S}^{(2)}_{lm,7}(\eta,\mathbf{x})&=&\frac{1}{4} h^{bc,(1)}\partial_l \partial_m h^{(1)}_{bc}  \ .
		\label{eq:7} 
	\end{eqnarray}
As shown in Eq.~(\ref{eq:1})--Eq.~(\ref{eq:7}), there are three kinds of source terms in Eq.~(\ref{eq:eq}). The source term $S_{lm,1}$ is the same as the source term of the second order SIGWs. The source terms $S_{lm,2}$ is composed of the product of the first order scalar perturbation $\phi^{(1)}$ and the first order tensor perturbation $h^{(1)}_{lm}$. The source terms $S_{lm,3} \sim S_{lm,7}$ are composed of the first order tensor perturbation $h^{(1)}_{lm}$.  In momentum space, Eq.~(\ref{eq:0}) can be written as 
\begin{equation}\label{eq:h}
	\begin{aligned}
		h^{\lambda,(2)''}(\eta,\mathbf{k})+2 \mathcal{H}h^{\lambda,(2)'}(\eta,\mathbf{k})+k^{2} h^{\lambda,(2)}(\eta,\mathbf{k}) = \sum^{7}_{i=1} 4\mathcal{S}_i^{\lambda,(2)}(\eta,\mathbf{k}) ~,
	\end{aligned}
\end{equation}
where $h^{\lambda,(2)}(\eta,\mathbf{k})=\varepsilon^{\lambda, ij}(\mathbf{k})h_{ij}^{(2)}(\eta,\mathbf{k})$ and  $S^{\lambda,(2)}(\eta,\mathbf{k})=-\varepsilon^{\lambda, lm}(\mathbf{k})S_{lm}^{(2)}(\eta,\mathbf{k})$. The $\varepsilon_{i j}^{\lambda}(\mathbf{k})$ is polarization tensor. The source terms $\mathcal{S}_i^{\lambda,(2)}(\eta,\mathbf{k})$ in Eq.~(\ref{eq:h}) can be expressed as
\begin{eqnarray}
	S_1^{\lambda,(2)}(\eta,\mathbf{k})&=&-\int\frac{d^3p}{(2\pi)^{3/2}}\varepsilon^{\lambda,lm}\left(\mathbf{k}\right)p_lp_mf^{(2)}_{1}\left( u,v,x \right)\Phi_{\mathbf{k}-\mathbf{p}}\Phi_{\mathbf{p}} \ ,
	\label{eq:s1}\\
	S_2^{\lambda,(2)}(\eta,\mathbf{k})&=&-\int\frac{d^3p}{(2\pi)^{3/2}}\varepsilon^{\lambda,lm}\left(\mathbf{k}\right) \varepsilon^{\lambda_1}_{lm}\left(\mathbf{p}\right)k^2 f^{(2)}_{2}\left( u,v,x \right) \Phi_{\mathbf{k}-\mathbf{p}}\mathbf{h}^{\lambda_1}_{\mathbf{p}} \ ,
	\label{eq:s2}\\
	S_3^{\lambda,(2)}(\eta,\mathbf{k})&=&-\int\frac{d^3p}{(2\pi)^{3/2}}\varepsilon^{\lambda,lm}\left(\mathbf{k}\right) \varepsilon^{\lambda_1,b}_{l}\left(\mathbf{k}-\mathbf{p}\right)\varepsilon^{\lambda_2}_{bm}\left(\mathbf{p}\right)k^2f^{(2)}_{3}\left(u,v,x\right)\mathbf{h}^{\lambda_1}_{\mathbf{k}-\mathbf{p}}\mathbf{h}^{\lambda_2}_{\mathbf{p}} \ ,
	\label{eq:s3}\\
	S_4^{\lambda,(2)}(\eta,\mathbf{k})&=&-\int\frac{d^3p}{(2\pi)^{3/2}}\varepsilon^{\lambda,lm}\left(\mathbf{k}\right) \varepsilon^{\lambda_1,bc}\left( \mathbf{k}-\mathbf{p} \right)\left(  2\varepsilon^{\lambda_2}_{mb}\left( \mathbf{p} \right)p_cp_l\right) f^{(2)}_4\left(u,v,x\right)\mathbf{h}^{\lambda_1}_{\mathbf{k}-\mathbf{p}}\mathbf{h}^{\lambda_2}_{\mathbf{p}} \ ,
	\label{eq:s4}\\
	S_5^{\lambda,(2)}(\eta,\mathbf{k})&=&-\int\frac{d^3p}{(2\pi)^{3/2}}\varepsilon^{\lambda,lm}\left(\mathbf{k}\right) \varepsilon^{\lambda_1}_{mc}\left(\mathbf{k}-\mathbf{p}\right)\varepsilon^{\lambda_2}_{lb}\left(\mathbf{p}\right)\left( k-p \right)^bp^c f^{(2)}_{5}\left(u,v,x\right)\mathbf{h}^{\lambda_1}_{\mathbf{k}-\mathbf{p}}\mathbf{h}^{\lambda_2}_{\mathbf{p}} \ ,
	\label{eq:s5}\\
	S_6^{\lambda,(2)}(\eta,\mathbf{k})&=&-\int\frac{d^3p}{(2\pi)^{3/2}}\varepsilon^{\lambda,lm}\left(\mathbf{k}\right) \varepsilon^{\lambda_1}_{bc}\left(\mathbf{k}-\mathbf{p}\right)\varepsilon^{\lambda_2}_{lm}\left(\mathbf{p}\right)p^bp^cf^{(2)}_6\left(u,v,x\right)\mathbf{h}^{\lambda_1}_{\mathbf{k}-\mathbf{p}}\mathbf{h}^{\lambda_2}_{\mathbf{p}} \ ,
	\label{eq:s6} \\
	S_7^{\lambda,(2)}(\eta,\mathbf{k})&=&-\int\frac{d^3p}{(2\pi)^{3/2}}\varepsilon^{\lambda,lm}\left(\mathbf{k}\right) \varepsilon^{\lambda_1,bc}\left(\mathbf{k}-\mathbf{p}\right)\varepsilon^{\lambda_2}_{bc}\left(\mathbf{p}\right)p_lp_m f^{(2)}_{7}\left(u,v,x\right) \mathbf{h}^{\lambda_1}_{\mathbf{k}-\mathbf{p}}\mathbf{h}^{\lambda_2}_{\mathbf{p}} \ ,
	\label{eq:s7}
\end{eqnarray}
where the transfer functions $f^{(2)}_i\left(u,v,x\right)$ are given by
\begin{eqnarray}
	f_1^{(2)}\left( u,v,x \right)&=&3T_{\phi}\left(ux \right)T_{\phi}\left( vx \right)+2ux \frac{d}{d(ux)}T_{\phi}\left( ux \right)T_{\phi}\left(vx \right)+x^2uv\frac{d}{d(ux)} T_{\phi}\left( ux \right)\frac{d}{d(vx)}T_{\phi}\left( vx \right) \ ,
	\label{eq:f1}\\
	f_2^{(2)}\left( u,v,x \right)&=&\frac{10u}{x}\frac{d}{d(ux)} T_{\phi}\left( ux \right)T_{h}\left( vx \right)+3u^2\frac{d^2}{d(ux)^2}T_{\phi}\left(ux \right)T_{h}\left( vx \right)+\frac{5}{3}u^2 T_{\phi}\left( ux \right)T_{h}\left( vx \right) \nonumber\\
	& &+\left(1-v^2-u^2\right)T_{\phi}\left( ux \right)T_{h}\left( vx \right)+2v^2T_{\phi}\left( ux \right)T_{h}\left( vx \right) \ ,
	\label{eq:f2}\\
	f_{3}^{(2)}\left(u,v,x\right)&=&-\frac{1-u^2-v^2}{4}T_{h}\left( ux \right)T_{h}\left( vx \right)-\frac{uv}{2}\frac{d}{d(ux)}T_{h}\left( ux \right)\frac{d}{d(ux)}T_{h}\left( vx \right) \ ,
	\label{eq:f3} \\
	f_i^{(2)}\left( u,v,x \right)&=&\frac{1}{2}T_{h}\left( ux \right)T_{h}\left( vx \right) \ , \  (i=4,5,6) \  , 
	\label{eq:f4} \\
	f_7^{(2)}\left( u,v,x \right)&=&\frac{1}{4}T_{h}\left( ux \right)T_{h}\left( vx \right) \ .
	\label{eq:f7}
\end{eqnarray}
Here, the explicit expressions of $T_{h}$ and $T_{\phi}$ have been given in Eq.~(\ref{eq:T}). We have defined $|k-p|=uk$ and $p=vk$.
Solving Eq.~(\ref{eq:h}) in terms of the Green's function method, we obtain
\begin{equation}\label{eq:h10}
	\begin{aligned}
		h^{\lambda,(2)}(\eta,\mathbf{k})=\sum^{7}_{i=1} h_i^{\lambda,(2)}(\eta,\mathbf{k}) \ ,
	\end{aligned}
\end{equation}
where $h_i^{\lambda,(2)}(\eta,\mathbf{k})$ $(i=1\sim 7)$ can be expressed as
\begin{eqnarray}
	h_1^{\lambda,(2)}(\eta,\mathbf{k})&=&-\int\frac{d^3p}{(2\pi)^{3/2}}\varepsilon^{\lambda,lm}\left(\mathbf{k}\right)p_lp_mI^{(2)}_{1}\left( u,v,x \right)\Phi_{\mathbf{k}-\mathbf{p}}\Phi_{\mathbf{p}} \ ,
	\label{eq:h1}\\
	h_2^{\lambda,(2)}(\eta,\mathbf{k})&=&-\int\frac{d^3p}{(2\pi)^{3/2}}\varepsilon^{\lambda,lm}\left(\mathbf{k}\right) \varepsilon^{\lambda_1}_{lm}\left(\mathbf{p}\right)k^2 I^{(2)}_{2}\left( u,v,x \right) \Phi_{\mathbf{k}-\mathbf{p}}\mathbf{h}^{\lambda_1}_{\mathbf{p}} \ ,
	\label{eq:h2}\\
	h_3^{\lambda,(2)}(\eta,\mathbf{k})&=&-\int\frac{d^3p}{(2\pi)^{3/2}}\varepsilon^{\lambda,lm}\left(\mathbf{k}\right) \varepsilon^{\lambda_1,b}_{l}\left(\mathbf{k}-\mathbf{p}\right)\varepsilon^{\lambda_2}_{bm}\left(\mathbf{p}\right)k^2I^{(2)}_{3}\left(u,v,x\right)\mathbf{h}^{\lambda_1}_{\mathbf{k}-\mathbf{p}}\mathbf{h}^{\lambda_2}_{\mathbf{p}} \ ,
	\label{eq:h3}\\
	h_4^{\lambda,(2)}(\eta,\mathbf{k})&=&-\int\frac{d^3p}{(2\pi)^{3/2}}\varepsilon^{\lambda,lm}\left(\mathbf{k}\right) \varepsilon^{\lambda_1,bc}\left( \mathbf{k}-\mathbf{p} \right)\left(  2\varepsilon^{\lambda_2}_{mb}\left( \mathbf{p} \right)p_cp_l\right) I^{(2)}_4\left(u,v,x\right)\mathbf{h}^{\lambda_1}_{\mathbf{k}-\mathbf{p}}\mathbf{h}^{\lambda_2}_{\mathbf{p}} \ ,
	\label{eq:h4}\\
	h_5^{\lambda,(2)}(\eta,\mathbf{k})&=&-\int\frac{d^3p}{(2\pi)^{3/2}}\varepsilon^{\lambda,lm}\left(\mathbf{k}\right) \varepsilon^{\lambda_1}_{mc}\left(\mathbf{k}-\mathbf{p}\right)\varepsilon^{\lambda_2}_{lb}\left(\mathbf{p}\right)\left( k-p \right)^bp^c I^{(2)}_{5}\left(u,v,x\right)\mathbf{h}^{\lambda_1}_{\mathbf{k}-\mathbf{p}}\mathbf{h}^{\lambda_2}_{\mathbf{p}} \ ,
	\label{eq:h5}\\
	h_6^{\lambda,(2)}(\eta,\mathbf{k})&=&-\int\frac{d^3p}{(2\pi)^{3/2}}\varepsilon^{\lambda,lm}\left(\mathbf{k}\right) \varepsilon^{\lambda_1}_{bc}\left(\mathbf{k}-\mathbf{p}\right)\varepsilon^{\lambda_2}_{lm}\left(\mathbf{p}\right)p^bp^cI^{(2)}_6\left(u,v,x\right)\mathbf{h}^{\lambda_1}_{\mathbf{k}-\mathbf{p}}\mathbf{h}^{\lambda_2}_{\mathbf{p}} \ ,
	\label{eq:h6} \\
	h_7^{\lambda,(2)}(\eta,\mathbf{k})&=&-\int\frac{d^3p}{(2\pi)^{3/2}}\varepsilon^{\lambda,lm}\left(\mathbf{k}\right) \varepsilon^{\lambda_1,bc}\left(\mathbf{k}-\mathbf{p}\right)\varepsilon^{\lambda_2}_{bc}\left(\mathbf{p}\right)p_lp_m I^{(2)}_{7}\left(u,v,x\right) \mathbf{h}^{\lambda_1}_{\mathbf{k}-\mathbf{p}}\mathbf{h}^{\lambda_2}_{\mathbf{p}} \ ,
	\label{eq:h7}
\end{eqnarray}
here the kernel functions $I^{(2)}_i\left( u,v,x \right)$ in Eq.~(\ref{eq:h1})--Eq.~(\ref{eq:h7}) are given by
\begin{equation}\label{eq:I}
	\begin{aligned}
		I^{(2)}_i\left( u,v,x \right)=\frac{4}{k^2} \int_{0}^{x} d\bar{x} \left( \frac{\bar{x}}{x}\sin\left( x-\bar{x} \right) f_i\left( u,v,\bar{x} \right)  \right) \ , \ (i=1 \sim 7) \ .
	\end{aligned}
\end{equation}
As we mentioned before, the source term $S_{lm,1}$ is the same as the source term of the second order SIGWs and therefore Eq.~(\ref{eq:h1}) is the formal expression of the second order SIGWs. The expressions in Eq.~(\ref{eq:h2})--Eq.~(\ref{eq:h7}) are quite different from the second order SIGWs in Eq.~(\ref{eq:h1}), we will study these expressions in Sec.~\ref{sec:3}.  In the end of this section, we calculate the kernel functions in Eq.~(\ref{eq:I}). We present the kernel functions  $\left(I_i(u=1,v=1,x)\right)^2$ $(i=1\sim 7)$ as function of $x=k\eta$ in Fig.~\ref{fig:kernel_function}. As shown in Fig.~\ref{fig:kernel_function}, the kernel function of second order SIGW $I_1^{(2)}$ is much larger than other kernel functions. The second largest kernel function is $I_2^{(2)}$, which is the kernel function of the source term $S^{(2),\lambda}_2\sim \phi^{(1)}h^{(1),\lambda}$.

\section{Energy density spectra of second order GWs}\label{sec:3}
\begin{figure}
	\includegraphics[scale = 0.8]{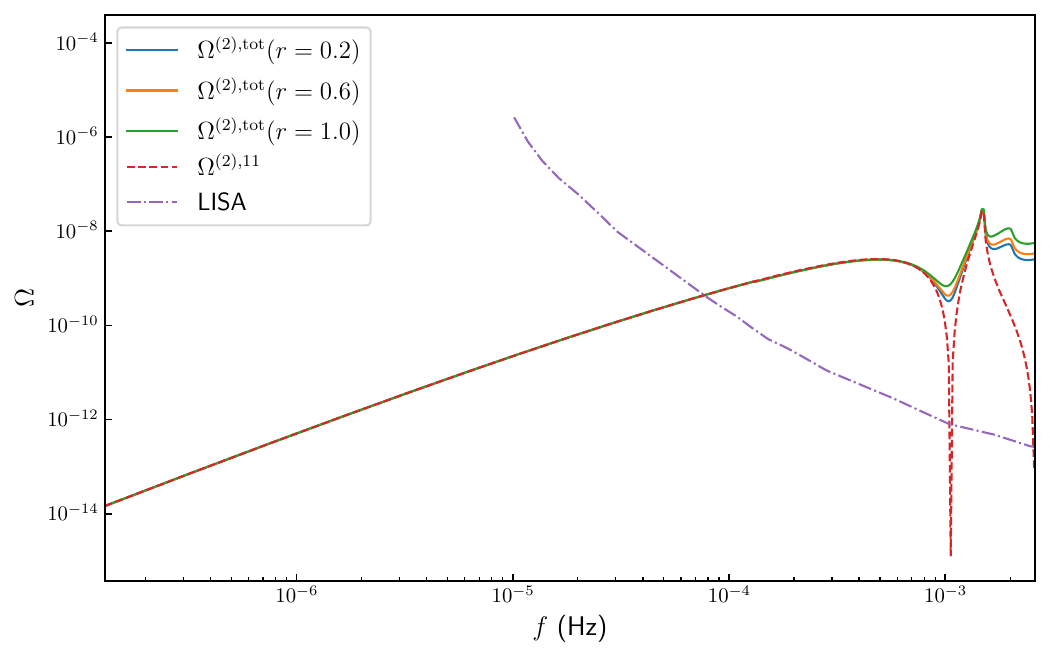}
	\caption{The total energy density spectra $\Omega^{(2), \mathrm{tot}}$ for different tensor-to-scalar ratio $r$ (solid curve) and the energy density spectrum for second order scalar induced gravitational waves $\Omega^{(2), 11}$ (orange dashed curve). The sensitivity curve of LISA \cite{Thrane:2013oya, Yuan:2019udt} (purple dotted curve) for $4$ years observation time is also shown. Here we have set the $A_\zeta=0.02$ and $f_*=1.3\times 10^{-3}$Hz. }\label{fig:Omega}
\end{figure}

The two-point function $\langle h^{\lambda,(2)} h^{\lambda',(2)}\rangle$ can be expressed as  
\begin{equation}\label{eq:sum}
	\begin{aligned}
\langle h^{\lambda,(2)} h^{\lambda',(2)}\rangle=\sum^{7}_{i,j=1}\langle h^{\lambda,(2)}_ih_j^{\lambda',(2)} \rangle=\langle h_1^{\lambda,(2)} h_1^{\lambda',(2)}\rangle+\langle h_1^{\lambda,(2)} h_2^{\lambda',(2)}\rangle+\cdot \cdot \cdot  \ .
	\end{aligned}
\end{equation}
The two-point correlation function and the power spectra are related by 
\begin{equation}\label{eq:P}
	\begin{aligned}
\langle h^{\lambda,(2)}(\eta,\mathbf{k}) h^{\lambda^{\prime},(2)}(\eta,\mathbf{k}')\rangle= \delta^{\lambda \lambda'}\delta\left(\mathbf{k}+\mathbf{k}'\right) \frac{2 \pi^{2}}{k^{3}} \mathcal{P}^{(2)}_{h}(\eta, \mathbf{k}) .
	\end{aligned}
\end{equation}
Substituting Eq.~(\ref{eq:h1})--Eq.~(\ref{eq:h7}) into Eq.~(\ref{eq:sum}), we will encounter six kinds of four-point functions of the primordial scalar and tensor perturbations, namely $\langle \Phi_{\mathbf{k}-\mathbf{p}}\mathbf{h}^{\lambda_1}_{\mathbf{p}} \mathbf{h}^{\lambda_1'}_{\mathbf{k}'-\mathbf{p}'}\mathbf{h}^{\lambda_2'}_{\mathbf{p}'} \rangle$, $\langle \Phi_{\mathbf{k}-\mathbf{p}}\Phi_{\mathbf{p}} \Phi_{\mathbf{k}'-\mathbf{p}'}\mathbf{h}^{\lambda_1'}_{\mathbf{p}'} \rangle$, $\langle \Phi_{\mathbf{k}-\mathbf{p}}\Phi_{\mathbf{p}} \mathbf{h}^{\lambda_1'}_{\mathbf{k}'-\mathbf{p}'}\mathbf{h}^{\lambda_2'}_{\mathbf{p}'} \rangle$, $\langle \Phi_{\mathbf{k}-\mathbf{p}}\Phi_{\mathbf{p}} \Phi_{\mathbf{k}'-\mathbf{p}'}\Phi_{\mathbf{p}'} \rangle$, $\langle \Phi_{\mathbf{k}-\mathbf{p}}\mathbf{h}^{\lambda_1}_{\mathbf{p}} \Phi_{\mathbf{k}'-\mathbf{p}'}\mathbf{h}^{\lambda_1'}_{\mathbf{p}'} \rangle$, and $\langle \mathbf{h}^{\lambda_1}_{\mathbf{k}-\mathbf{p}}\mathbf{h}^{\lambda_2}_{\mathbf{p}} \mathbf{h}^{\lambda_1'}_{\mathbf{k}'-\mathbf{p}'}\mathbf{h}^{\lambda_2'}_{\mathbf{p}'} \rangle$. These four-point functions can be studied in terms of the Wick’s theorem. We assume that the two-point function $\langle \Phi_{\mathbf{k}_1}\mathbf{h}^{\lambda}_{\mathbf{k}_2} \rangle$=0 for arbitrary $\mathbf{k}_1$ and $\mathbf{k}_2$. Therefore, we only need to consider last three four-point functions. The explicit expressions of the four-point functions are given in Appendix.~\ref{sec:A}. Substituting Eq.~(\ref{eq:h1}) and Eq.~(\ref{eq:A1})--Eq.~(\ref{eq:A3}) into Eq.~(\ref{eq:P}), we obtain 
\begin{eqnarray}
		\mathcal{P}^{(2),11}_h&=&\frac{1}{4}\int_{0}^{\infty} dv\int_{|1-v|}^{|1+v|}du \left( \frac{ 4v^2-\left( 1+v^2-u^2 \right)^2}{4uv} \right)^2 \left( k^2 I^{(2)}_1\left( u,v,x \right)  \right)^2 \mathcal{P}^{(1)}_{\Phi}\left( ku\right) \mathcal{P}^{(1)}_{\Phi}\left( vk \right)\ ,
		\label{eq:P11} \\
		\mathcal{P}^{(2),22}_h&=&\frac{1}{4}\int_{0}^{\infty} dv\int_{|1-v|}^{|1+v|}du ~ \frac{1}{64(uv)^2} \left(\frac{16 \left(-u^2+v^2+1\right)^2}{v^2}+\left(\frac{\left(-u^2+v^2+1\right)^2}{v^2}+4\right)^2\right) \nonumber\\
	  	& &\times \left( k^2 I^{(2)}_2\left( u,v,x \right)  \right)^2 \mathcal{P}^{(1)}_{\Phi}\left( ku\right) \mathcal{P}_{h}^{(1)}\left( vk \right) \ ,
	  	\label{eq:P22} \\
	  	\mathcal{P}^{(2),ij}_h&=&\frac{1}{4}\int_{0}^{\infty} dv\int_{|1-v|}^{|1+v|}du~\frac{\mathbb{P}^{ij}\left( u,v \right)}{\left(uv\right)^2}  \left( k^2 I^{(2)}_i\left( u,v,x \right)  k^2I^{(2)}_j\left( u,v,x \right)  \right) \nonumber\\
	  	& &\times \mathcal{P}_{h}^{(1)}\left( ku\right) \mathcal{P}_{h}^{(1)}\left( vk \right) \ , \  \left(i,j=3\sim 7\right) \ ,
	  	\label{eq:Pij} 
\end{eqnarray}
where $\mathcal{P}^{(1)}_{\Phi}$ and $\mathcal{P}^{(1)}_{h}$ are primordial power spectra of $\Phi_{\mathbf{k}}$ and $\mathbf{h}_{\mathbf{k}}$, respectively. In Eq.~(\ref{eq:Pij}), the polynomials $\mathbb{P}^{ij}\left( u,v \right)$ $\left(i,j=3\sim 7\right)$ can be calculated in terms of the contraction of the polarization tensor $\varepsilon^{\lambda}_{ij}\left(\mathbf{k} \right)$. The polarization tensors are given in Appendix.~\ref{sec:B} and the explicit expressions of $\mathbb{P}^{ij}\left( u,v \right)$ $\left(i,j=3\sim 7\right)$ are given in Appendix.~\ref{sec:C}. The power spectrum $\mathcal{P}^{11}_h$ is the power spectrum of second order scalar induced gravitational waves, it comes from the source term $S^{(2)}_{lm,1}$. The power spectrum $\mathcal{P}^{22}_h$ comes from the source term $S^{(2)}_{lm,2}$ which is the source term of the first order scalar perturbation and the first order tensor perturbation $\phi^{(1)}h^{(1),\lambda}$. The power spectra $\mathcal{P}^{ij}_h$ $\left(i,j=3\sim 7\right)$ come from the source terms $S^{(2)}_{lm,3}\sim S^{(2)}_{lm,7}$ which are the source terms of the first order tensor perturbation $h^{(1),\lambda_1}h^{(1),\lambda_2}$. The total energy density spectra of second order GWs is defined as \cite{Maggiore:1999vm}
\begin{equation}\label{eq:Omega}
	\begin{aligned}
		\Omega_{\mathrm{GW}}^{(2)}(\eta, k)=\frac{\rho^{(2)}_{\mathrm{GW}}(\eta, k)}{\rho_{\mathrm{tot}}(\eta)}=\frac{1}{24}\left(\frac{k}{a(\eta) H(\eta)}\right)^{2} {\mathcal{P}^{(2)}_{h}(\eta, k)} \ ,
	\end{aligned}
\end{equation}
where
\begin{equation}\label{eq:P12}
	\begin{aligned}
		{\mathcal{P}^{(2)}_{h}(\eta, k)}=\mathcal{P}^{(2),11}_{h}(\eta, k)+{\mathcal{P}^{(2),22}_{h}(\eta, k)}+\sum_{i,j=3}^{7} {\mathcal{P}^{(2),ij}_{h}(\eta, k)} \ .
	\end{aligned}
\end{equation}
In Eq.~(\ref{eq:P12}), $\mathcal{P}^{(1)}_h$ is the power spectrum of primordial GWs. The last three terms are the power spectra of second order GWs induced by primordial scalar and tensor perturbations. 
Here we consider the monochromatic primordial power spectra, namely
\begin{equation}
	\begin{aligned}
	 \mathcal{P}_{\zeta}=A_{\zeta}k_*\delta\left( k-k_* \right) \ , \ \mathcal{P}_{h}=A_{h}k_*\delta\left( k-k_* \right) \ ,
	\end{aligned}
\end{equation}
where $k_*$ is the wavenumber at which the power spectrum has a $\delta$ function peak. As we mentioned before, $\mathcal{P}_{\zeta}=9/4\mathcal{P}_{\Phi}$  is the  primordial power spectra of $\zeta_{\mathbf{k}}=3/2\Phi_{\mathbf{k}}$.  As mentioned in Sec.~\ref{sec:1}, the large primordial tensor perturbations on small scales may be realized by many models of early universe. The large tensor perturbations on small scales with peaks may be constructed in these framework with fine-tuning \cite{Nakama:2016enz,Nakama:2015nea,Kobayashi:2011nu,Boyle:2003km,Copeland:2008kz}. In Fig.~\ref{fig:Omega}, we plot the current energy density spectra of GWs
\begin{equation}\label{eq:tot_spectrum}
	\begin{aligned}
		\Omega_{\mathrm{GW}}(\eta_0, k)\simeq \Omega_r \times \Omega_{\mathrm{GW}}(\eta, k) \ ,
	\end{aligned}
\end{equation}
where $\Omega_r$ is the density parameter of radiation at present. In Fig.~\ref{fig:Omega}, the red dashed curve represents the power spectrum of second order GWs induced by  $\mathcal{P}_{\zeta}=A_{\zeta}k_*\delta\left( k-k_* \right)$, which was first studied in Ref.~\cite{Bugaev:2010bb,Saito:2009jt}. The solid curves represent the total power spectrum of second order GWs for different tensor-to-scalar ratio $r$. It shows that the effects of the first order tensor perturbation enhance the density spectrum significantly for high frequency second order gravitational waves. 

\begin{figure}
	\includegraphics[scale = 0.8]{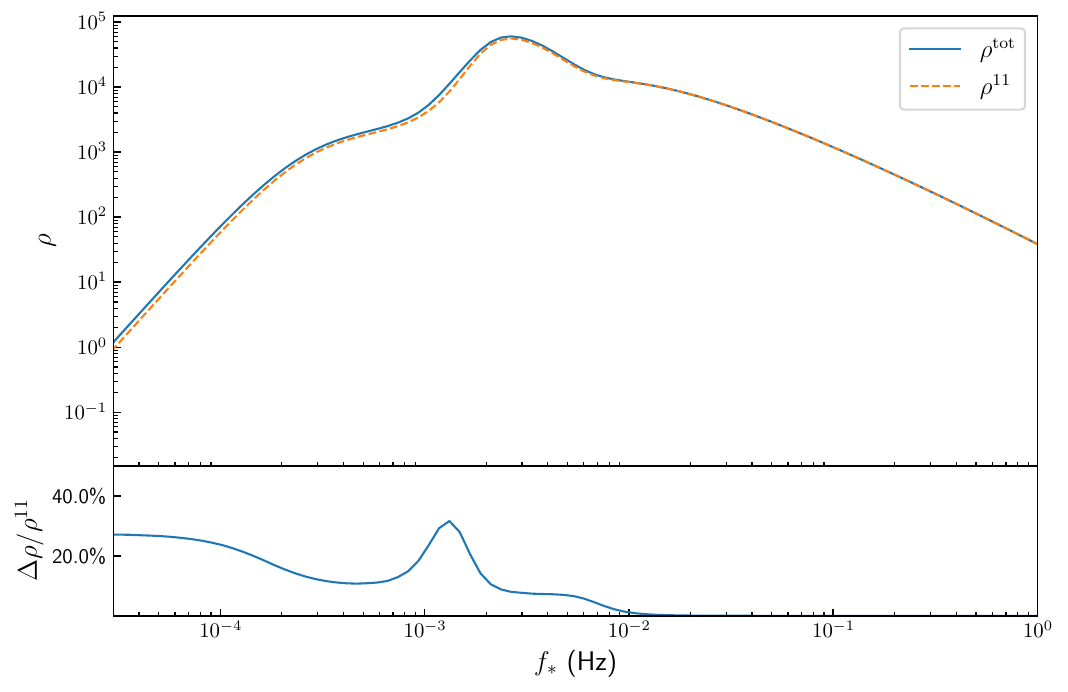}
	\caption{The SNR of LISA for $\Omega^{(2),\mathrm{tot}}$ ($\rho^\mathrm{tot}$, blue solid curve) and $\Omega^{(2),11}$ ($\rho^\mathrm{11}$, red dashed curve), where $f_*=k_*/(2\pi)$. We also give $\Delta \rho/\rho^{11} = \rho^\mathrm{tot}/\rho^{11}-1$ (bottom panel). }\label{fig:SNR_LISA}
\end{figure}

\begin{figure}
	\includegraphics[scale = 0.8]{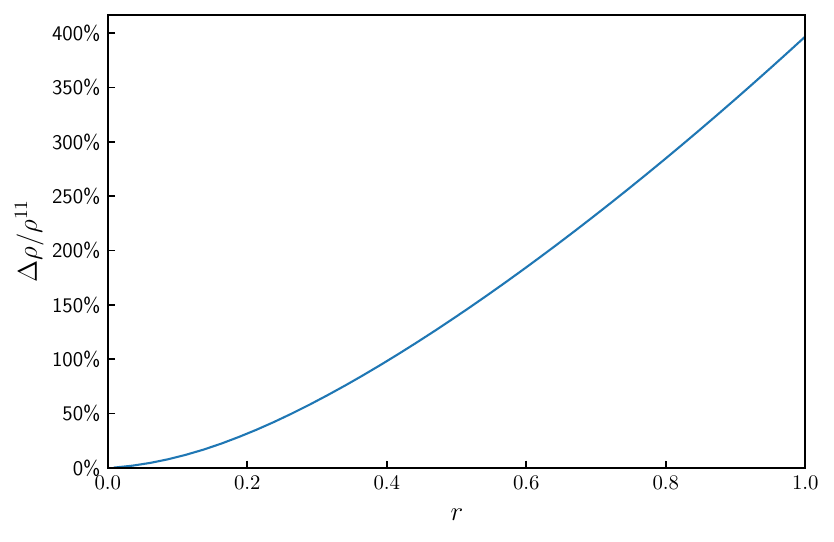}
	\caption{$\Delta \rho/\rho^{11} = \rho^\mathrm{tot}/\rho^{11}-1$ for different tensor-to-scalar ratio $r$. We have set $f_*=1.3\times 10 ^{-3}$. }\label{fig:SNR2}
\end{figure}

To quantify the effect of the first order tensor perturbation to the induced GWs, we give the SNR $\rho$ in Fig.~\ref{fig:SNR_LISA}. For LISA, SNR is given by \cite{Yuan:2019udt}
\begin{equation}
	\rho = \sqrt{T} \left[ \int df \left( \frac{\Omega_\mathrm{GW}(f)}{\Omega_n(f)} \right)^2\right]^{1/2} \ ,
\end{equation}
where $\Omega_n(f) = 2\pi^2f^3S_n/3H_0^2$ and $S_n$ is the strain noise power spectral density, $T$ is the observation time. In this paper, we set $T=4$ yr. Other explicit parameters of LISA can be found in Ref.~\cite{Robson:2018ifk}. As shown in Fig.~\ref{fig:SNR_LISA}, the effects of the primordial tensor perturbation lead to an around $30\%$ increase of the SNR for LISA observations for the tensor-to-scalar ratio $r=A_{h}/A_{\zeta}=0.2$. And it is found that the source term $S_2 \sim \phi^{(1)}h^{(1)}$ dominates the effects of the first order tensor perturbation. The effects of the source term $S_i \sim h^{(1)}h^{(1)}$$(i=3 \sim 7)$ are negligible for $r=0.2$. Moreover, as presented in Fig.~\ref{fig:SNR2}, we calculate SNR for LISA observations for different tensor-to-scalar ratio $r$. It shows that $\Delta \rho/\rho^{11}>100\%$ if $r>0.4$,namely, the effect of primordial tensor perturbation will be larger than the effect of primordial scalar perturbation if $r>0.4$.

\section{Conclusion and Discussion}\label{sec:4}
The scalar induced gravitational waves have been studied for many years. However, the first order tensor perturbation has been always neglected. In this paper, we consider the effects of the first order tensor perturbation, the first order scalar and tensor perturbations all induce the second order GWs. We conclude that the effects of the first order tensor perturbation enhance the density spectrum significantly for high frequency second order GWs even for small tensor-to-scalar ratio $r$. For $f_*=1.3\times 10^{-3}$ and $A_{\zeta}=0.02$, the effects of the primordial tensor perturbation will lead to an around $100\%$ increase of the signal-to-noise ratio (SNR) for LISA observations if $r>0.4$.

The explicit expressions of the power spectra of the second order GWs were presented. Here, we only considered the monochromatic primordial power spectra, one can calculate the power spectra of the second order GWs for various primordial power spectra in terms of Eq.~(\ref{eq:P11})--Eq.~(\ref{eq:Pij}). As we mentioned in Sec.\ref{sec:1}, the large primordial tensor perturbations on small scales may be realized by many models of early universe, these models can be constrained by the current and future observations of second order GWs.

The second order power spectra of GWs induced by primordial scalar and tensor perturbations can be written as $\mathcal{P}^{(2)}_h \sim \langle h^{(2)}h^{(2)} \rangle +\langle h^{(1)}h^{(3)}  \rangle $. Here, we only studied the contributions of second order GWs. We found that the  effects of the primordial tensor perturbation in $\langle h^{(2)}h^{(2)}  \rangle$ are important in the UV region ($k> k_*$). In Ref.~\cite{Chen:2022dah}, the IR behaviors ($k\ll k_*$) of $\langle h^{(1)}h^{(3)}  \rangle$ were investigated in detail. Further researches might be given in the future.

Our results show that the effects of the first order tensor perturbation on small scales are very important for the second order GWs. Therefore, it is necessary to consider the effects of the first order tensor perturbation when one calculates the higher order scalar and vector perturbations \cite{Inomata:2020cck,Lu:2008ju}. The first order tensor perturbation on small scales will affect the observations related to second order scalar, vector, and tensor perturbations. Meanwhile, these second order perturbations will affect the the observations of higher order perturbations such as the third order induced GWs \cite{Zhou:2021vcw}.  Perhaps, a complete study on these higher order perturbations might be presented in the future.\cite{Chang:2022vlv}

\begin{acknowledgements} 
	We thank Prof.~S. Wang, Dr.~Y.H. Yu, and Dr.~ J.P. Li for the useful discussions. This work has been funded by the National Nature Science Foundation of China under grant No. 12075249 and 11690022, and the Key Research Program of the Chinese Academy of Sciences under Grant No. XDPB15.
\end{acknowledgements}

\appendix
\section{Four-point function of primordial perturbations}\label{sec:A}
In this appendix, we calculate three kinds of four-point functions used in Sec.~\ref{sec:3} in terms of the Wick’s theorem.
\begin{equation} \label{eq:A1}
	\begin{aligned}
		\left\langle\Phi_{\mathbf{k}-\mathbf{p}} \Phi_{\mathbf{p}} \Phi_{\mathbf{k}'-\mathbf{p}'} \Phi_{\mathbf{p}'}\right\rangle=&\left\langle\Phi_{\mathbf{k}-\mathbf{p}} \Phi_{\mathbf{k}'-\mathbf{p}'}\right\rangle\left\langle\Phi_{\mathbf{p}} \Phi_{\mathbf{p}'}\right\rangle+\left\langle\Phi_{\mathbf{k}-\mathbf{p}} \Phi_{\mathbf{p}'}\right\rangle\left\langle\Phi_{\mathbf{p}} \Phi_{\mathbf{k}'-\mathbf{p}'}\right\rangle \\
		=&\left(\delta\left(\mathbf{k}-\mathbf{p}+\mathbf{k}'-\mathbf{p}'\right)\frac{2\pi^2}{|\mathbf{k}-\mathbf{p}|^3}\mathcal{P}_{\Phi}(|\mathbf{k}-\mathbf{p}|) \right)\left(\delta\left(\mathbf{p}+\mathbf{p}'\right)\frac{2\pi^2}{p^3}\mathcal{P}_{\Phi}(p) \right) \\
		+&\left(\delta\left(\mathbf{k}-\mathbf{p}+\mathbf{p}'\right)\frac{2\pi^2}{|\mathbf{k}-\mathbf{p}|^3}\mathcal{P}_{\Phi}(|\mathbf{k}-\mathbf{p}|) \right)\left(\delta\left(\mathbf{k}'-\mathbf{p}'+\mathbf{p}\right)\frac{2\pi^2}{p^3}\mathcal{P}_{\Phi}(p) \right) \\
		=&\left(\delta\left(\mathbf{k}+\mathbf{k}'\right)\frac{2\pi^2}{|\mathbf{k}-\mathbf{p}|^3}\mathcal{P}_{\Phi}(|\mathbf{k}-\mathbf{p}|) \right)\left(\delta\left(\mathbf{p}+\mathbf{p}'\right)\frac{2\pi^2}{p^3}\mathcal{P}_{\Phi}(p) \right) \\
		+&\left(\delta\left(\mathbf{k}+\mathbf{k}'\right)\frac{2\pi^2}{|\mathbf{k}-\mathbf{p}|^3}\mathcal{P}_{\Phi}(k_1) \right)\left(\delta\left(\mathbf{k}'-\mathbf{p}'+\mathbf{p}\right)\frac{2\pi^2}{p^3}\mathcal{P}_{\Phi}(p) \right) \\
		=&\delta\left(\mathbf{k}+\mathbf{k}'\right) \frac{(2\pi^2)^2}{p^3|\mathbf{k}-\mathbf{p}|^3} \left(\delta\left(\mathbf{p}+\mathbf{p}'\right)+\delta\left(\mathbf{k}'-\mathbf{p}'+\mathbf{p}\right) \right)\mathcal{P}_{\Phi}(|\mathbf{k}-\mathbf{p}|)\mathcal{P}_{\Phi}(p) \ ,
	\end{aligned}
\end{equation}

\begin{equation}\label{eq:A2}
	\begin{aligned}
		\langle \mathbf{h}_{\mathbf{k}-\mathbf{p}}^{\lambda_1}\mathbf{h}_{\mathbf{p}}^{\lambda_2} \mathbf{h}_{\mathbf{k}'-\mathbf{p}'}^{\lambda_1'}\mathbf{h}_{\mathbf{p}'}^{\lambda_2'} \rangle&=\langle \mathbf{h}_{\mathbf{k}-\mathbf{p}}^{\lambda_1} \mathbf{h}_{\mathbf{k}'-\mathbf{p}'}^{\lambda_1'}  \rangle\langle \mathbf{h}_{\mathbf{p}}^{\lambda_2}\mathbf{h}_{\mathbf{p}'}^{\lambda_2'} \rangle+\langle \mathbf{h}_{\mathbf{k}-\mathbf{p}}^{\lambda_1}\mathbf{h}_{\mathbf{p}'}^{\lambda_2'} \rangle \langle \mathbf{h}_{\mathbf{p}}^{\lambda_2} \mathbf{h}_{\mathbf{k}'-\mathbf{p}'}^{\lambda_1'} \rangle \\
		&=\left(\delta^{\lambda_1\lambda'_1} \delta\left(\mathbf{k}-\mathbf{p}+\mathbf{k}'-\mathbf{p}'\right)\frac{2\pi^2}{|\mathbf{k}-\mathbf{p}|^3}\mathcal{P}_{h}(|\mathbf{k}-\mathbf{p}|) \right)\left(\delta^{\lambda_2\lambda'_2} \delta\left(\mathbf{p}+\mathbf{p}'\right)\frac{2\pi^2}{p^3}\mathcal{P}_{h}(p) \right) \\
		&+\left(\delta^{\lambda_1\lambda'_2} \delta\left(\mathbf{k}-\mathbf{p}+\mathbf{p}'\right)\frac{2\pi^2}{|\mathbf{k}-\mathbf{p}|^3}\mathcal{P}_{h}(|\mathbf{k}-\mathbf{p}|) \right)\left(\delta^{\lambda_2\lambda'_1} \delta\left(\mathbf{p}+\mathbf{k}'-\mathbf{p}'\right)\frac{2\pi^2}{p^3}\mathcal{P}_{h}(p) \right) \\
		&=\delta\left(\mathbf{k}+\mathbf{k}'\right) \frac{(2\pi^2)^2}{p^3|\mathbf{k}-\mathbf{p}|^3} \left(\delta^{\lambda_1\lambda'_1}\delta^{\lambda_2\lambda'_2}\delta\left(\mathbf{p}+\mathbf{p}'\right)+\delta^{\lambda_1\lambda'_2}\delta^{\lambda_2\lambda'_1}\delta\left(\mathbf{k}'-\mathbf{p}'+\mathbf{p}\right) \right)\mathcal{P}_{h}(|\mathbf{k}-\mathbf{p}|)\mathcal{P}_{h}(p) \ ,
	\end{aligned}
\end{equation}

\begin{equation} \label{eq:A3}
	\begin{aligned}
		\langle \Phi_{\mathbf{k}-\mathbf{p}}\mathbf{h}_{\mathbf{p}}^{\lambda_1} \Phi_{\mathbf{k}'-\mathbf{p}'}\mathbf{h}_{\mathbf{p}'}^{\lambda'_1}  \rangle =&\langle \Phi_{\mathbf{k}-\mathbf{p}} \Phi_{\mathbf{k}'-\mathbf{p}'}\rangle\langle \mathbf{h}_{\mathbf{p}}^{\lambda_1}\mathbf{h}_{\mathbf{p}'}^{\lambda'_1}  \rangle \\
		=&\left(\delta\left(\mathbf{k}+\mathbf{k}'\right)\frac{2\pi^2}{|\mathbf{k}-\mathbf{p}|^3}\mathcal{P}_{\Phi}(|\mathbf{k}-\mathbf{p}|) \right)\left(\delta^{\lambda_1\lambda'_1} \delta\left(\mathbf{p}+\mathbf{p}'\right)\frac{2\pi^2}{p^3}\mathcal{P}_{h}(p) \right) \\
		=&\delta\left(\mathbf{k}+\mathbf{k}'\right)\delta^{\lambda_1\lambda'_1} \frac{(2\pi^2)^2}{p^3|\mathbf{k}-\mathbf{p}|^3} \delta\left(\mathbf{p}+\mathbf{p}'\right)\mathcal{P}_{\Phi}(|\mathbf{k}-\mathbf{p}|)\mathcal{P}_{h}(p) \ .
	\end{aligned}
\end{equation}

\section{Polarization tensor}\label{sec:B}
In this appendix, we present the explicit expressions of the polarization tensor for a given coordinate system. The polarization tensor is defined as
\begin{equation}
	\begin{aligned}
		\varepsilon^{\times}_{ij}\left(\mathbf{k}  \right)=\frac{1}{\sqrt{2}}\left( e_i\left( \mathbf{k} \right)\bar{e}_j\left( \mathbf{k} \right)+\bar{e}_i\left( \mathbf{k} \right)e_j\left( \mathbf{k} \right)  \right) \ ,
	\end{aligned} 
\end{equation}
\begin{equation}
	\begin{aligned}
		\varepsilon^{+}_{ij}\left(\mathbf{k}  \right)=\frac{1}{\sqrt{2}}\left( e_i\left( \mathbf{k} \right)e_j\left( \mathbf{k} \right)-\bar{e}_i\left( \mathbf{k} \right)\bar{e}_j\left( \mathbf{k} \right)  \right) \ ,
	\end{aligned} 
\end{equation}
where $\left(\mathbf{k}_i/|k|,e_i\left( \mathbf{k} \right),\bar{e}_i\left( \mathbf{k} \right)  \right)$ is the normalized bases in three dimensional momentum space. For a given coordinate system, we set
\begin{equation}
	\begin{aligned}
		\mathbf{k}=\left(0,0,k  \right) \ , \ e_i\left( \mathbf{k} \right)=\left( 1,0,0 \right) \ , \ \bar{e}_i\left( \mathbf{k} \right)=\left( 0,1,0 \right) \ .
	\end{aligned} 
\end{equation}
Then the polarization tensors $	\varepsilon^{\lambda}_{ij}\left(\mathbf{k}-\mathbf{p}  \right)$ and $	\varepsilon^{\lambda}_{ij}\left(\mathbf{p}  \right)$ can be written as
\begin{equation}
	\begin{aligned}
		\varepsilon^{\times}_{ij}\left(\mathbf{k}-\mathbf{p}  \right)&=\frac{1}{\sqrt{2}}\left( e_i\left( \mathbf{k}-\mathbf{p} \right)\bar{e}_j\left( \mathbf{k}-\mathbf{p} \right)+\bar{e}_i\left( \mathbf{k}-\mathbf{p} \right)e_j\left( \mathbf{k}-\mathbf{p} \right)  \right) \ , \\
		\varepsilon^{+}_{ij}\left(\mathbf{k}-\mathbf{p}  \right)&=\frac{1}{\sqrt{2}}\left( e_i\left( \mathbf{k}-\mathbf{p} \right)e_j\left( \mathbf{k}-\mathbf{p} \right)-\bar{e}_i\left( \mathbf{k}-\mathbf{p} \right)\bar{e}_j\left( \mathbf{k}-\mathbf{p} \right)  \right) \ , \\
		\varepsilon^{\times}_{ij}\left(\mathbf{p}  \right)&=\frac{1}{\sqrt{2}}\left( e_i\left( \mathbf{p} \right)\bar{e}_j\left( \mathbf{p} \right)+\bar{e}_i\left( \mathbf{p} \right)e_j\left( \mathbf{p} \right)  \right)\ , \\ \varepsilon^{+}_{ij}\left(\mathbf{p}  \right)&=\frac{1}{\sqrt{2}}\left( e_i\left( \mathbf{p} \right)e_j\left( \mathbf{p} \right)-\bar{e}_i\left( \mathbf{p} \right)\bar{e}_j\left( \mathbf{p} \right)  \right) \ ,
	\end{aligned} 
\end{equation}
where 
\begin{equation}
	\begin{aligned}
		\mathbf{k}-\mathbf{p}&=k\left(-\sqrt{ v^2-\frac{1}{4}  \left(-u^2+v^2+1\right)^2},0,\frac{1}{2} \left(u^2-v^2+1\right)  \right)  \ , \\
		e_i\left( \mathbf{k}-\mathbf{p} \right)&=\left(\frac{u^2-v^2+1}{2 u},0,\frac{\sqrt{-u^4+2 u^2 v^2+2 u^2-v^4+2 v^2-1}}{2 u}  \right)\ , \\
		\bar{e}_i\left( \mathbf{k}-\mathbf{p} \right)&=\left( 0,1,0 \right) \ ,
	\end{aligned} 
\end{equation}
\begin{equation}
	\begin{aligned}
		\mathbf{p}&=k\left(\sqrt{ v^2-\frac{1}{4} \left(-u^2+v^2+1\right)^2},0,\frac{1}{2}  \left(-u^2+v^2+1\right) \right)  \ , \\
		e_i\left( \mathbf{p} \right)&=\left(-\frac{-u^2+v^2+1}{2 v},0,\frac{\sqrt{-u^4+2 u^2 \left(v^2+1\right)-\left(v^2-1\right)^2}}{2 v}  \right)\ , \\
		\bar{e}_i\left( \mathbf{p} \right)&=\left( 0,1,0 \right) \ .
	\end{aligned} 
\end{equation}

\section{$\mathbb{P}^{ij}\left( u,v \right)$}\label{sec:C}

The explicit expressions of $\mathbb{P}^{ij}$, $(i, j = 3\sim 7)$ are given by
\begin{equation}
	\begin{split}
		\mathbb{P}^{33} = \frac{1}{256 u^4 v^4}\left(u^4-2 u^2 \left(v^2+1\right)+\left(v^2-1\right)^2\right)^2 \left(u^4+6 u^2 \left(v^2+1\right)+v^4+6 v^2+1\right) \ ,
	\end{split}
\end{equation}

\begin{equation}
	\begin{split}
		\mathbb{P}^{44} = \frac{1}{256 u^4 v^4} &\left(u^4-2 u^2 \left(v^2+1\right)+\left(v^2-1\right)^2\right)^2 \times \\
		&\left(u^6-u^4 \left(5 v^2+3\right)+u^2 \left(-25 v^4+6 v^2+3\right)-\left(v^2-1\right)^2 \left(3 v^2+1\right)\right) \ ,
	\end{split}
\end{equation}

\begin{equation}
	\begin{split}
		\mathbb{P}^{55} = &\frac{1}{1024 u^4 v^4} \left(u^4-2 u^2 \left(v^2+1\right)+\left(v^2-1\right)^2\right)^2 \times \\
		&\left(u^8-4 u^6 \left(v^2-1\right)+2 u^4 \left(3 v^4-2 v^2-5\right)-4 u^2 \left(v^6+v^4-9 v^2-1\right)+\left(v^2-1\right)^2 \left(v^4+6 v^2+1\right)\right)\ ,
	\end{split}
\end{equation}

\begin{equation}
	\begin{split}
		\mathbb{P}^{66} = &\frac{1}{1024 u^4 v^4} \left(u^4-2 u^2 \left(v^2+1\right)+\left(v^2-1\right)^2\right)^2 \\
		&\left(u^8-4 u^6 v^2+2 u^4 \left(3 v^4+8 v^2-1\right)-4 u^2 v^2 \left(v^4+8 v^2+3\right)+\left(v^2+1\right)^2 \left(v^4+14 v^2+1\right)\right) \ ,
	\end{split}
\end{equation}

\begin{equation}
	\begin{split}
		\mathbb{P}^{77} = &\frac{1}{1024 u^4 v^4}\left(u^4-2 u^2 \left(v^2+1\right)+\left(v^2-1\right)^2\right)^2 \times \\
		&\left(u^8+4 u^6 \left(7 v^2-1\right)+u^4 \left(70 v^4-60 v^2+6\right)+4 u^2 \left(v^2-1\right)^2 \left(7 v^2-1\right)+\left(v^2-1\right)^4\right) \ ,
	\end{split}
\end{equation}

\begin{equation}
	\begin{split}
		\mathbb{P}^{34} = \frac{1}{512 u^4 v^4}\left(u^4-2 u^2 \left(v^2+1\right)+\left(v^2-1\right)^2\right)^2 \left(3 u^4+2 u^2 \left(9 v^2-1\right)+3 v^4-2 v^2-1\right) \ ,
	\end{split}
\end{equation}

\begin{equation}
	\begin{split}
		\mathbb{P}^{35} = \frac{1}{512 u^4 v^4}\left(u^4-2 u^2 \left(v^2+1\right)+\left(v^2-1\right)^2\right)^2 \left(u^6-u^4 \left(v^2-5\right)-u^2 \left(v^4+18 v^2+5\right)+v^6+5 v^4-5 v^2-1\right) \ ,
	\end{split}
\end{equation}

\begin{equation}
	\begin{split}
		\mathbb{P}^{36} = \frac{1}{512 u^4 v^4}\left(u^4-2 u^2 \left(v^2+1\right)+\left(v^2-1\right)^2\right)^2 \left(u^6-u^4 \left(v^2-7\right)-u^2 \left(v^4+6 v^2-7\right)+v^6+7 v^4+7 v^2+1\right) \ ,
	\end{split}
\end{equation}

\begin{equation}
	\begin{split}
		\mathbb{P}^{37} = \frac{1}{512 u^4 v^4}\left(u^4-2 u^2 \left(v^2+1\right)+\left(v^2-1\right)^2\right)^2 \left(u^6+u^4 \left(15 v^2-1\right)+u^2 \left(15 v^4-6 v^2-1\right)+\left(v^2-1\right)^2 \left(v^2+1\right)\right) \ ,
	\end{split}
\end{equation}

\begin{equation}
	\begin{split}
		\mathbb{P}^{45} = \frac{1}{512 u^4 v^4}\left(u^4-v^4-6 v^2-1\right) \left(u^4-2 u^2 \left(v^2+1\right)+\left(v^2-1\right)^2\right)^3 \ ,
	\end{split}
\end{equation}

\begin{equation}
	\begin{split}
		\mathbb{P}^{46} = &\frac{1}{512 u^4 v^4}\left(u^4-2 u^2 \left(v^2+1\right)+\left(v^2-1\right)^2\right)^2 \times \\
		&\left(u^8-2 u^6 v^2+2 u^4 \left(v^2-1\right)+2 u^2 v^2 \left(v^4+2 v^2-11\right)-v^8-6 v^6+6 v^2+1\right) \ ,
	\end{split}
\end{equation}

\begin{equation}
	\begin{split}
		\mathbb{P}^{47} = \frac{1}{512 u^4 v^4} &\left(u^2-v^2-1\right) \left(u^4-2 u^2 \left(v^2+1\right)+\left(v^2-1\right)^2\right)^2 \times \\
		&\left(u^6+3 u^4 \left(5 v^2-1\right)+3 u^2 \left(5 v^4-6 v^2+1\right)+\left(v^2-1\right)^3\right) \ ,
	\end{split}
\end{equation}

\begin{equation}
	\begin{split}
		\mathbb{P}^{56} = \frac{1}{1024 u^4 v^4}\left(u^2-v^2-1\right) \left(u^2-v^2+1\right) \left(u^4-2 u^2 \left(v^2-3\right)+v^4+6 v^2+1\right) \left(u^4-2 u^2 \left(v^2+1\right)+\left(v^2-1\right)^2\right)^2 \ ,
	\end{split}
\end{equation}

\begin{equation}
	\begin{split}
		\mathbb{P}^{57} = \frac{1}{1024 u^4 v^4}\left(u^4+6 u^2 v^2+v^4-1\right) \left(u^4-2 u^2 \left(v^2+1\right)+\left(v^2-1\right)^2\right)^3 \ ,
	\end{split}
\end{equation}

\begin{equation}
	\begin{split}
		\mathbb{P}^{67} = &\frac{1}{1024 u^4 v^4}\left(u^4-2 u^2 \left(v^2+1\right)+\left(v^2-1\right)^2\right)^2 \times \\
		&\left(u^8+4 u^6 \left(v^2-1\right)-2 u^4 \left(5 v^4+2 v^2-3\right)+4 u^2 \left(v^6+9 v^4-v^2-1\right)+\left(v^2-1\right)^2 \left(v^4+6 v^2+1\right)\right) \ .
	\end{split}
\end{equation}

\bibliography{biblio}

\end{document}